\documentclass{ws-procs961x669}            

\providecommand{\sorthelp}[1]{}
\let\vec\mathbf
\def\kms{\ifmmode $\,km\,s$^{-1}\else \,km\,s$^{-1}$\fi}
\def\pdeg{\ifmmode $\setbox0=\hbox{$^{\circ}$}\rlap{\hskip.11\wd0 .}$^{\circ}
          \else \setbox0=\hbox{$^{\circ}$}\rlap{\hskip.11\wd0 .}$^{\circ}$\fi}

\def\smicano{{\tt SMICA-NOSZ}}

\def\milca{{\tt MILCA}}
\def\twodilc{{\tt 2D-ILC}}
\def\Planck{\textit{Planck}}
\def\healpix{\texttt{HEALPix}}

\def\figsubcap#1{\par\noindent\centering\footnotesize #1}
\def\arcm{\ifmmode {^{\scriptstyle\prime}}
          \else $^{\scriptstyle\prime}$\fi}

\usepackage{soul}
\usepackage{epstopdf}
\begin{document}

\title{The CMB Dipole: Eppur Si Muove}

\author{R. M. Sullivan$^*$, D. Scott}

\address{Physics and Astronomy, University of British Columbia,\\
Vancouver, BC, Canada\\
$^*$E-mail: rsullivan@phas.ubc.ca\\
www.ubc.ca}

\begin{abstract}
The largest temperature anisotropy in the cosmic microwave background (CMB) is the dipole. The simplest interpretation of the dipole is that it is due to our motion with respect to the rest frame of the CMB. As well as creating the $\ell$=1 mode of the CMB sky, this motion affects all astrophysical observations by modulating and aberrating sources across the sky. It can be seen in galaxy clustering, and in principle its time derivative through a dipole-shaped acceleration pattern in quasar positions. Additionally, the dipole modulates the CMB temperature anisotropies with the same frequency dependence as the thermal Sunyaev-Zeldovich (tSZ) effect and so these modulated CMB anisotropies can be extracted from the tSZ maps produced by \Planck. Unfortunately this measurement cannot determine if the dipole is due to our motion, but it does provide an independent measure of the dipole and a validation of the $y$~maps. This measurement, and a description of the first-order terms of the CMB dipole, are outlined here.
\end{abstract}

\keywords{Cosmic Microwave Background; Cosmic Microwave Background dipole; Special relativity; Thermal Sunyaev-Zeldovich effect.}

\bodymatter
\section{The CMB Sky from \Planck}
\label{sec:CMBPlanck}
\Planck\footnote{\Planck\ (\url{http://www.esa.int/Planck}) is a project of the European Space Agency (ESA) with instruments provided by two scientific consortia funded by ESA member states and led by Principal Investigators from France and Italy, telescope reflectors provided through a collaboration between ESA and a scientific consortium led and funded by Denmark, and additional contributions from NASA (USA).}  was a space-based telescope that measured the microwave sky in nine wavebands, allowing it to capture not only the cosmic microwave background (CMB) but also several Galactic and extragalactic foreground components. This is most clearly seen in figure~51 from Ref.~\citenum{planck2014-a12}, which shows the various wavebands of the \Planck\ satellite and the frequency spectra of the foreground signals across those bands. One signal of interest to this study is the thermal Sunyaev-Zeldovich (tSZ) effect, which produces so-called $y$-type distortion signals. This comes from CMB photons being inverse-Compton scattered, mostly through hot galaxy clusters, which makes holes (or lowers the flux) at low frequencies and up-scatters (or makes an excess flux) at high frequencies. This signal allows us to construct a novel and independent measure of the CMB dipole because temperature anisotropies stemming from the CMB dipole contaminate the $y$~maps. It can also be used as a valuable test of the quality of the $y$~maps. We will start in Sec.~\ref{sec:unboostCMB} by setting out relevant notation for the unboosted CMB sky. Next, in Sec.~\ref{sec:boostCMB} we will boost the CMB sky, and explore the relevant terms that arise from that boost in the subsections. Of particular relevance, in Sec.~\ref{sec:tszandcmb} we will discuss our measurement of the dipole modulation terms that mix with the tSZ effect. We will finish in Sec.~\ref{sec:concl} with a short discussion and conclusion regarding our work. 

\section{The Unboosted CMB sky}
\label{sec:unboostCMB}
To derive the connection between the $y$~map and the dipole we will begin by defining some useful terms regarding the unboosted CMB sky: 

\begin{align}
    x&\equiv\frac{h\nu}{k_\mathrm{B}T};\\
    I&\equiv\frac{2k^3_\mathrm{B}T^3}{h^2c^2}\frac{x^3}{e^x-1};\\
    f(x)&\equiv\frac{xe^x}{e^x-1};\\
    Y(x)&\equiv x\frac{e^x+1}{e^x-1}-4.
\end{align}
 These are the dimensionless frequency, the standard Planck blackbody intensity function, the frequency dependence of the CMB anisotropies and the relative frequency dependence of the tSZ effect or $y$~type distortions, respectively. 
 Thus, to first order the anisotropies of intensity measured by \Planck\ can be written as
 \begin{align}
     \frac{\delta I(\hat{\vec{n}})}{If(x)}=\frac{\delta T(\hat{\vec{n}})}{T_{\rm CMB}}+y(\hat{\vec{n}})Y(x),
     \label{equ:CMBsky}
 \end{align}
where $\hat{\vec{n}}$ is the line of sight direction on the sky and we have only considered the temperature anisotropies and the $y$~signals here. 

\section{The Boosted CMB sky}
\label{sec:boostCMB}
If we apply a boost to Eq.~\ref{equ:CMBsky}, with a dimensionless velocity $\bm{\beta}$, we find
\begin{align}
    \frac{\delta I'(\vec{\hat{n}'})}{If(x)}=&\beta\mu+\frac{\delta T(\vec{\hat{n}'})}{T_{\rm CMB}}(1+3\beta\mu)\nonumber\\
    &+Y(x)\left(y(\vec{\hat{n}'})(1+3\beta\mu)+\beta\mu\frac{\delta T(\vec{\hat{n}'})}{T_{\rm CMB}}\right)\nonumber\\
    &+\beta\mu y(\vec{\hat{n}}')\left(Y^2(x)-x\frac{dY(x)}{dx}\right)+\mathcal{O}(\beta^2),
    \label{equ:boostCMB} 
\end{align}
where $\mu=\cos(\theta)$, and $\theta$ is the angle between the boost $\bm{\beta}$ and the line of sight $\vec{\hat{n}}'$. 
The first line has the same frequency dependence as thermal fluctuations and so appear in typical CMB temperature anisotropy maps. Crucially for our analysis, the middle line has the same frequency dependence as $y$-type distortions and thus describes the signals in the $y$~map. The final line has more obscure frequency dependence and is not discussed here. Additionally, the direction of the incoming photons will change from $\vec{\hat{n}}$ to $\vec{\hat{n}'}$, where $\vec{\hat{n}'}=\vec{\hat{n}}-\nabla(\vec{\hat{n}}\cdot\vec{\beta})$; this deflection of the photons by $\nabla(\vec{\hat{n}}\cdot\vec{\beta})$ is due to aberration, an effect that is not unique to the microwave sky and occurs for all astronomical observations. We will now discuss each of these terms in turn.

\subsection{The CMB dipole: $\beta\mu$}
\label{sec:cmbdip}
\begin{figure}[h]
\begin{center}
\includegraphics[width=5in]{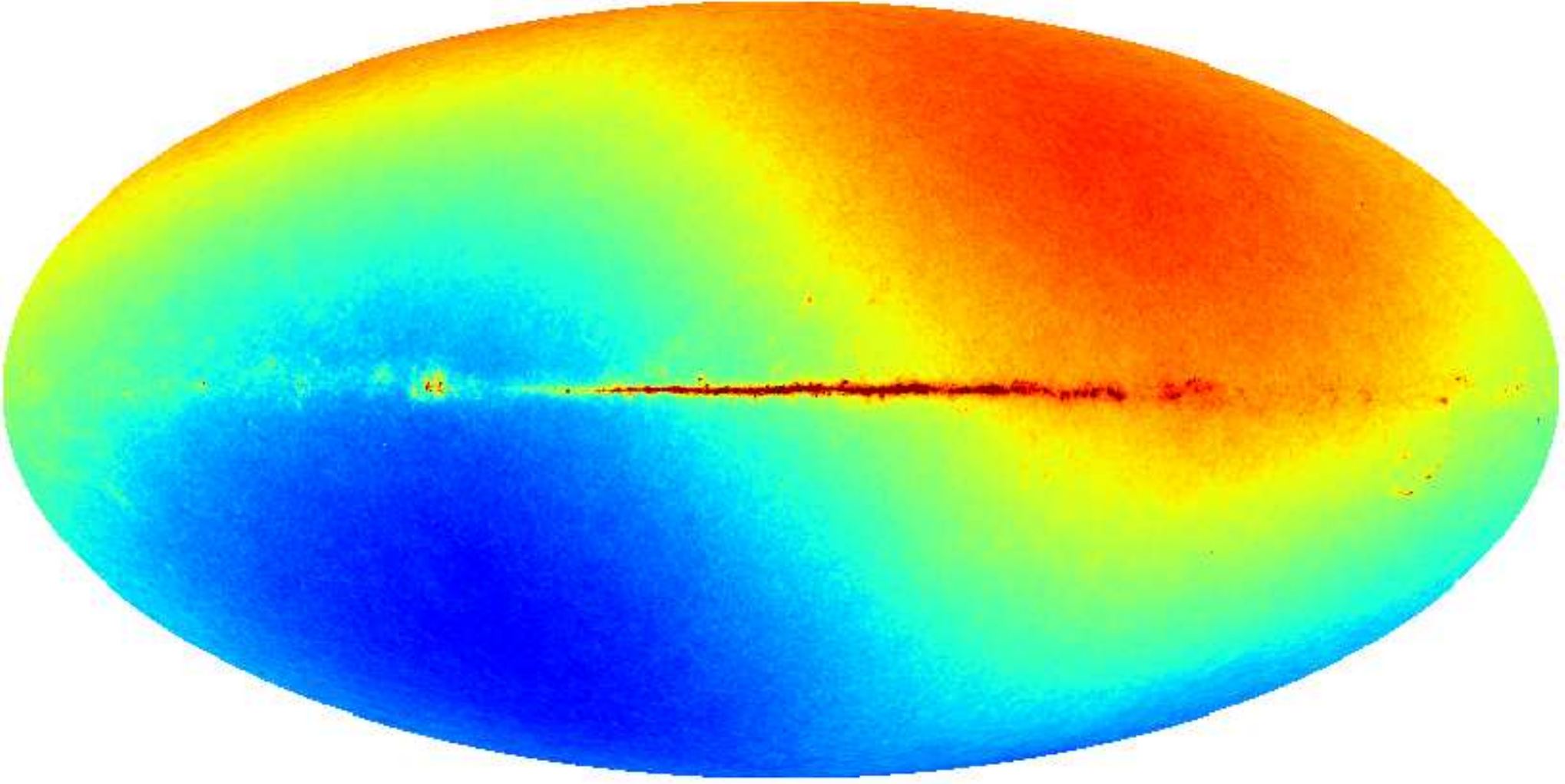}
\end{center}
\caption{\Planck\ 100-GHz channel map from the NPIPE (PR4) data release \citep{planck2020-LVII}, showing the dominant $\ell=1$ mode or dipole across the sky. The temperature difference across the sky here is 3.36\,mK.}
\label{fig:cmbdip}
\end{figure}

In the first line of Eq.~\ref{equ:boostCMB}, the term $\beta\mu$ describes the pure CMB dipole, as discussed previously. This mainly (or perhaps entirely) comes from our local motion with respect to the CMB rest frame and it has been
previously measured in Refs.~\citenum{Kogut1993}, \citenum{Fixsen1996}, and \citenum{hinshaw2009}, and most recently in Refs.~\citenum{planck2016-l01}, \citenum{planck2016-l02}, and
\citenum{planck2016-l03}. Taking the
large dipole as being solely caused by our motion, the velocity is $v = (369.82 \pm
0.11) \kms$ in the direction $(l, b) = (264\pdeg021 \pm 0\pdeg011, 48\pdeg253 \pm
0\pdeg005)$ \citep{planck2016-l01} and can be easily seen in the CMB frequency maps, such as in Fig.~\ref{fig:cmbdip}. 

\subsection{Aberration and Modulation of the CMB anisotropies: $(1+3\beta\mu)\delta T(\vec{\hat{n}'})/T_{\rm CMB}$}
\label{sec:cmbabandmod}
\begin{figure}
    \centering
    \parbox{3in}{\includegraphics[width=\hsize]{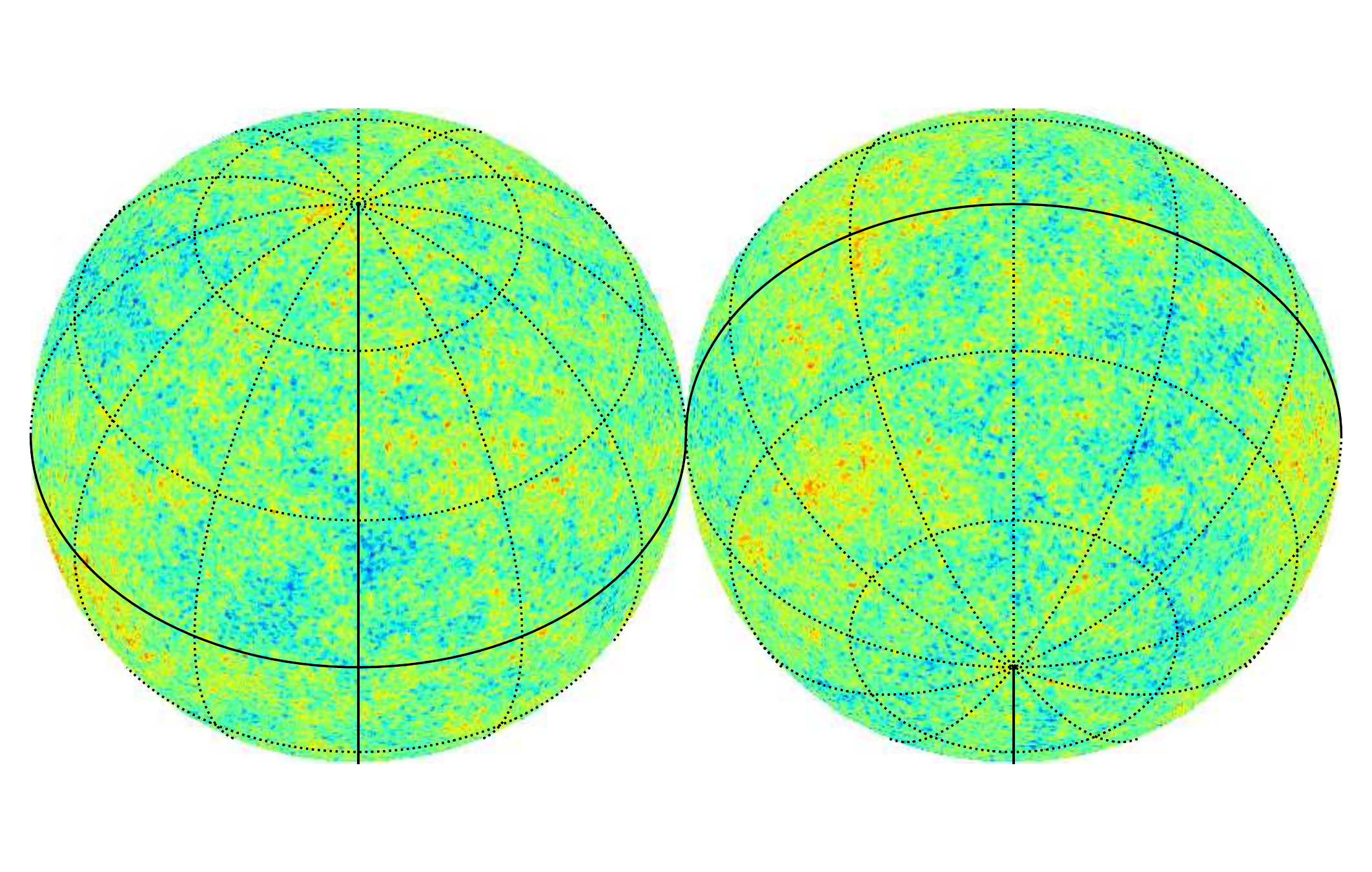}\figsubcap{(a) The unboosted CMB sky.}}
    \parbox{3in}{\includegraphics[width=\hsize]{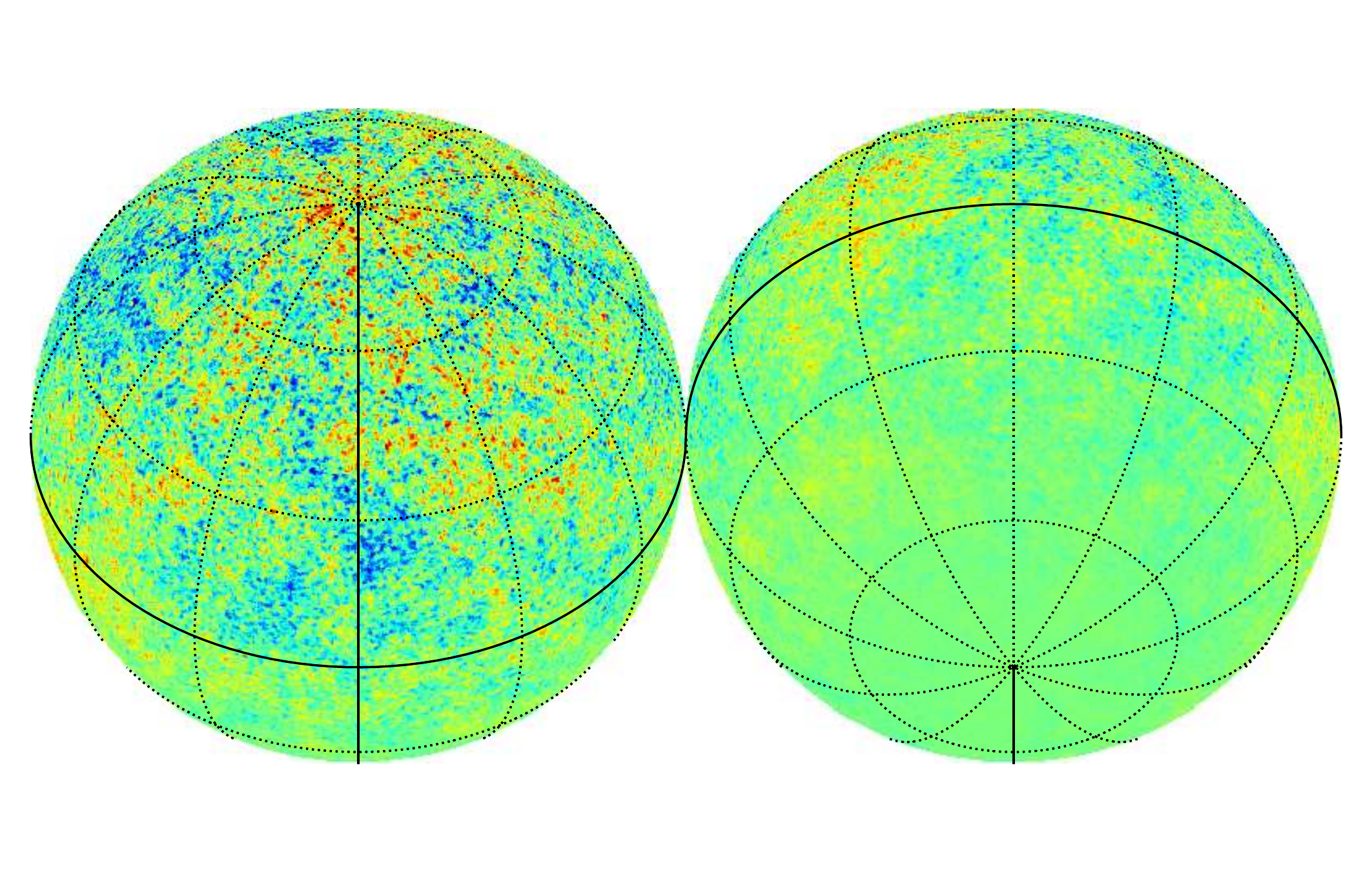}\figsubcap{(b) The modulated CMB sky.}}
    \parbox{3in}{\includegraphics[width=\hsize]{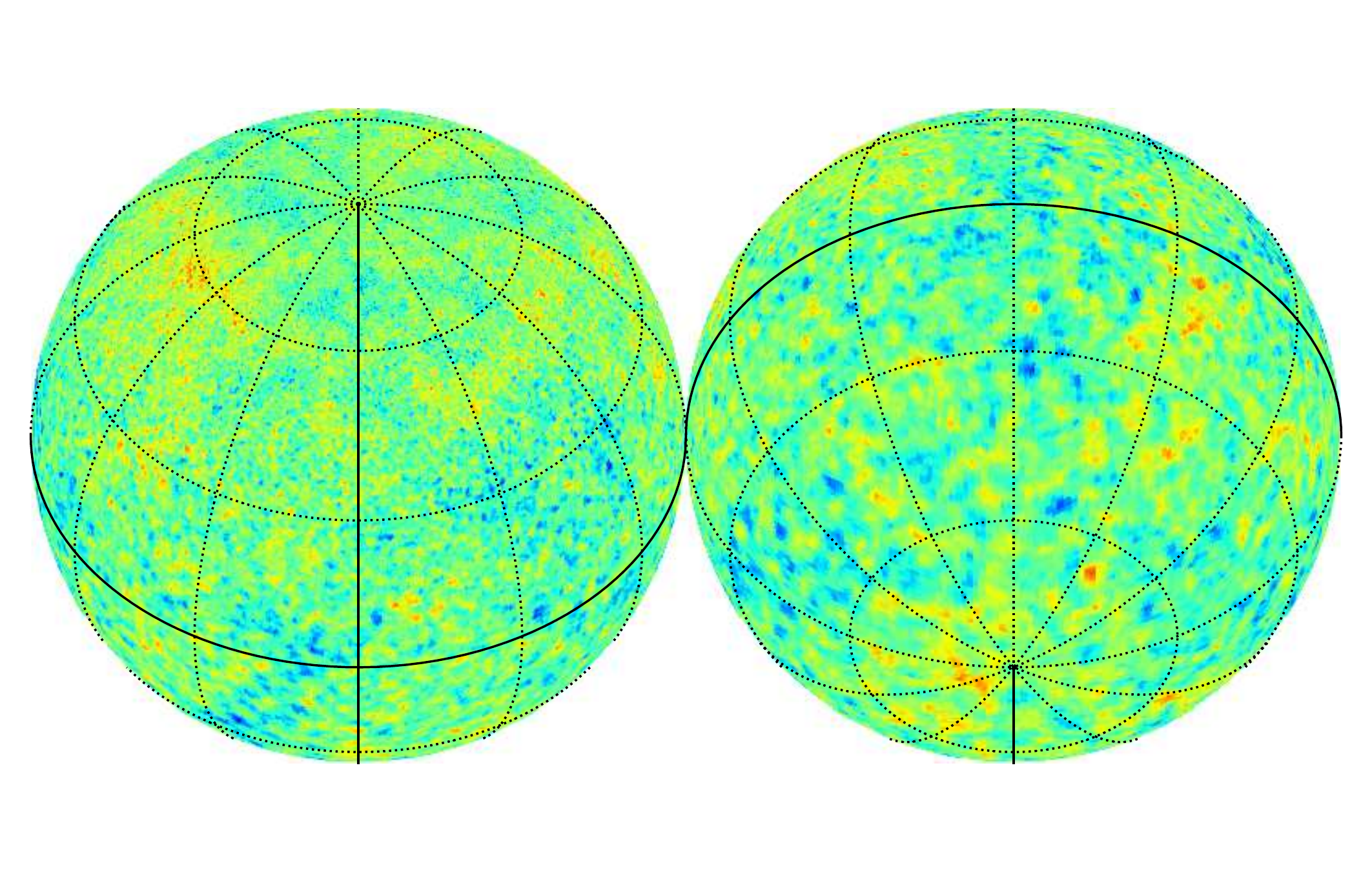}\figsubcap{(c) The aberrated CMB sky.}}
    \caption{Here the CMB sky is shown unboosted in (a), with a modulation from a boost of 90\,\% the speed of light in (b), and with aberration from a boost of 90\,\% of the speed of light in (c). In the case of modulation, the anisotropies are more intense in the forward direction and less so in the reverse direction, whereas the aberration condenses the anisotropies in the forwards direction and causes them to be more spread out in the reverse direction.
    }
    \label{fig:abmod}
\end{figure}

The second term in the first line of Eq.~\ref{equ:boostCMB} is the dipole aberration and modulation of the temperature anisotropies of the CMB. The modulation causes the temperature anisotropies to be brighter in the forwards direction, and dimmer in the reverse direction. The aberration causes the anisotropies to be more condensed in the forwards direction, and more stretched out in the reverse direction (effectively the same as $\ell=1$ lensing). These two effects can be seen in Fig.~\ref{fig:abmod}. This effect was first measured in Ref.~\citenum{planck2013-pipaberration} to about 4$\,\sigma$.

\subsection{Temperature modulation and the tSZ effect: $Y(x)\beta\mu\delta T(\vec{\hat{n}'})/T_{\rm CMB}$}
\label{sec:tszandcmb}
The second line of Eq.~\ref{equ:boostCMB} shows the dipole-generated signals in the $y$~maps produced by \Planck. The first half is the same modulation and aberration terms as were seen in the temperature anisotropies; however, the final term is due to the second-order expansion of the intensity about $T_{\rm CMB}$ and adds a contribution to the $y$~maps from the temperature anisotropies. 

We can look for this signal by cross-correlating a template map, derived
from the CMB temperature data, with a $y$~map.
To this end, we use the so-called \twodilc\ CMB temperature map which was produced by the
``Constrained ILC'' component-separation method designed by
Ref.~\citenum{Remazeilles2011} to explicitly null out the contribution from the
$y$-type spectral distortions in the CMB map. We also use the \smicano\ temperature map, similarly produced with the express intent of removing the $y$-type spectral distortions, and which was generated for the \Planck\ 2018 data release \citep{planck2016-l04}.
Likewise, we use the corresponding \twodilc\ $y$~map, and the \Planck\ \milca\ $y$~map, which explicitly null out the contributions from a (differential) blackbody spectral distribution in the $y$~map \citep{Hurier2013, planck2014-a28}.
If we multiply our CMB map with
$\beta\mu$ and cross-correlate that with our tSZ map, then
we can directly probe the dipole modulation.

\begin{figure}[htbp!]
\begin{center}
\parbox{2.1in}{\includegraphics[width=\hsize]{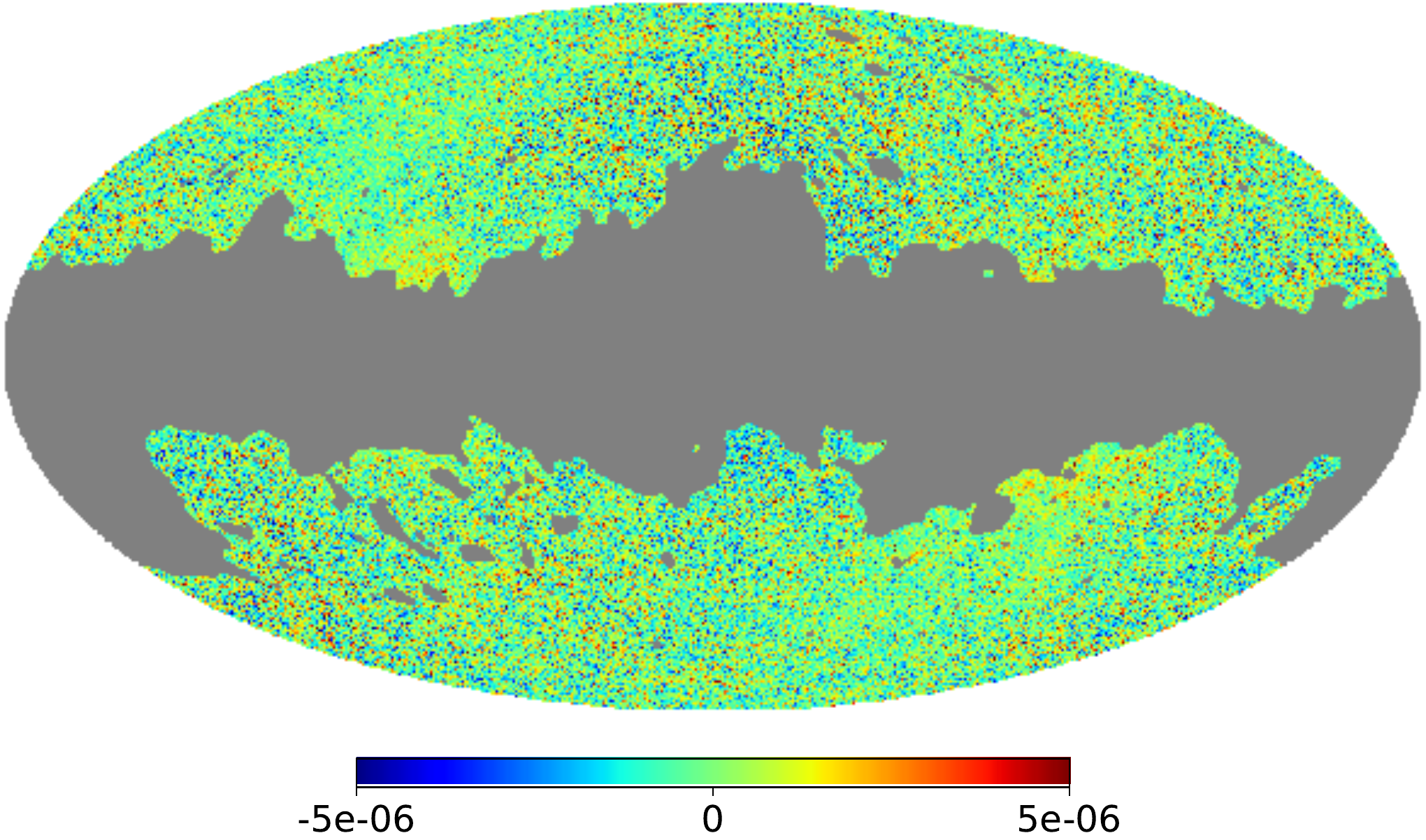}}
\parbox{2.1in}{\includegraphics[width=\hsize]{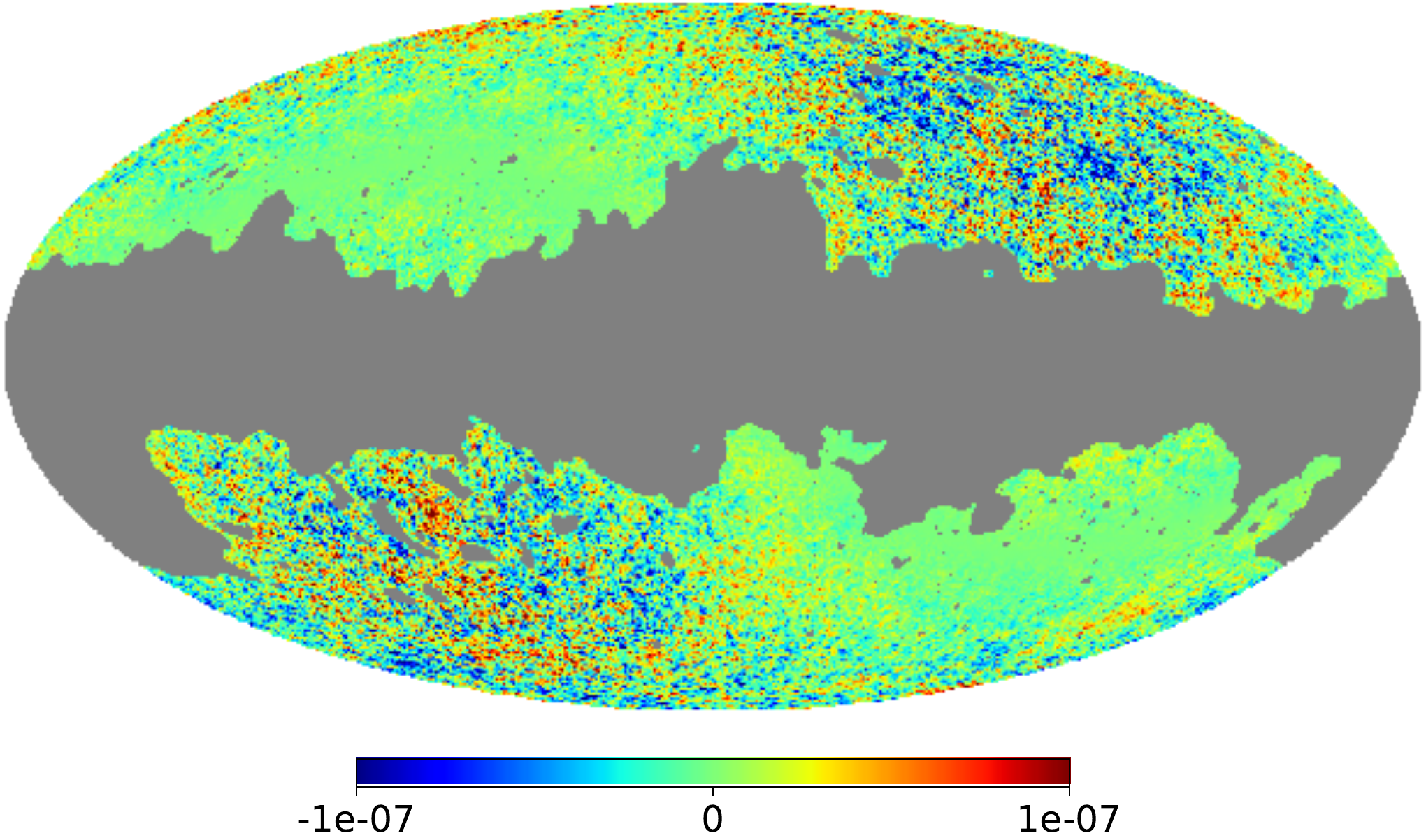}}
\parbox{2.1in}{\includegraphics[width=\hsize]{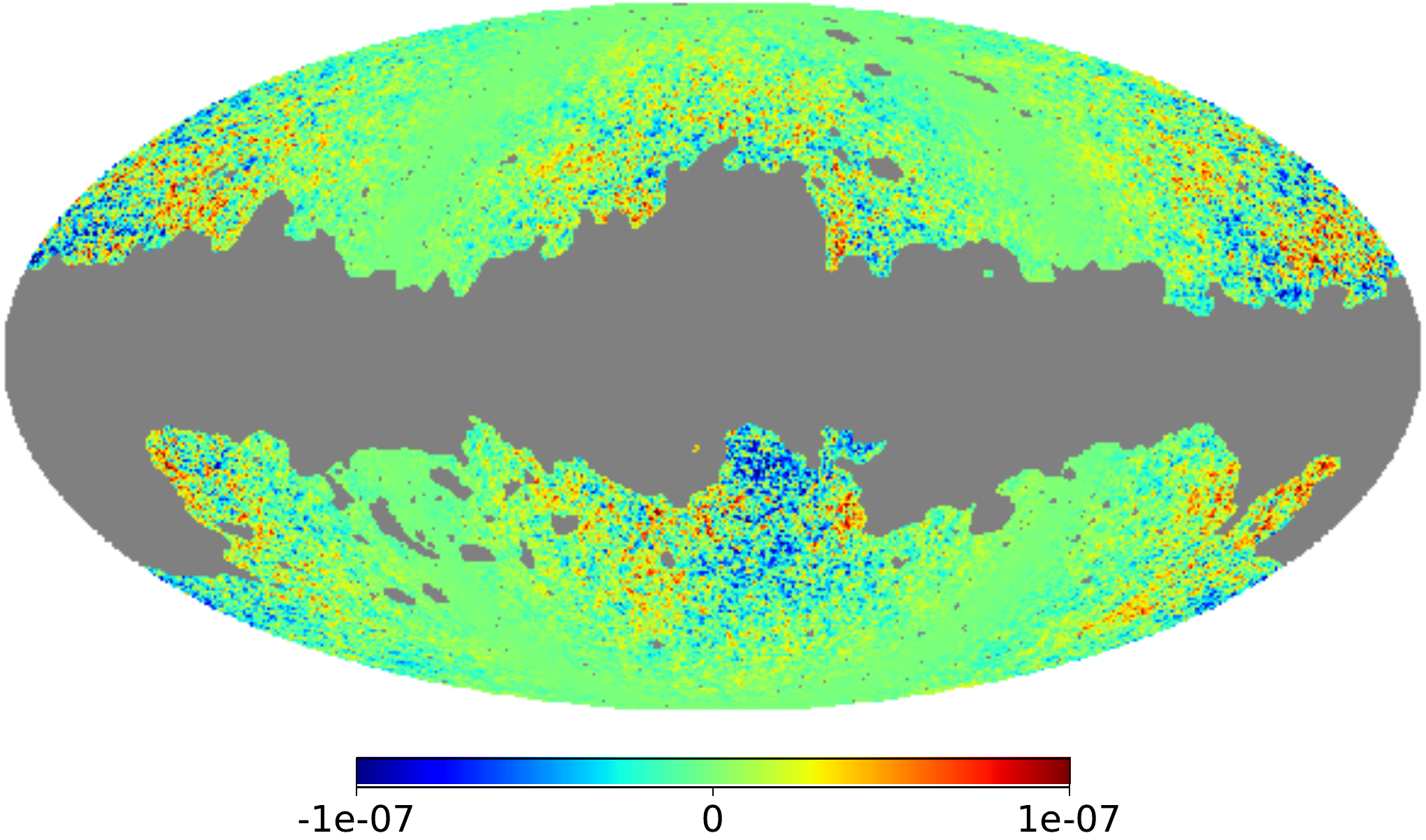}}
\parbox{2.1in}{\includegraphics[width=\hsize]{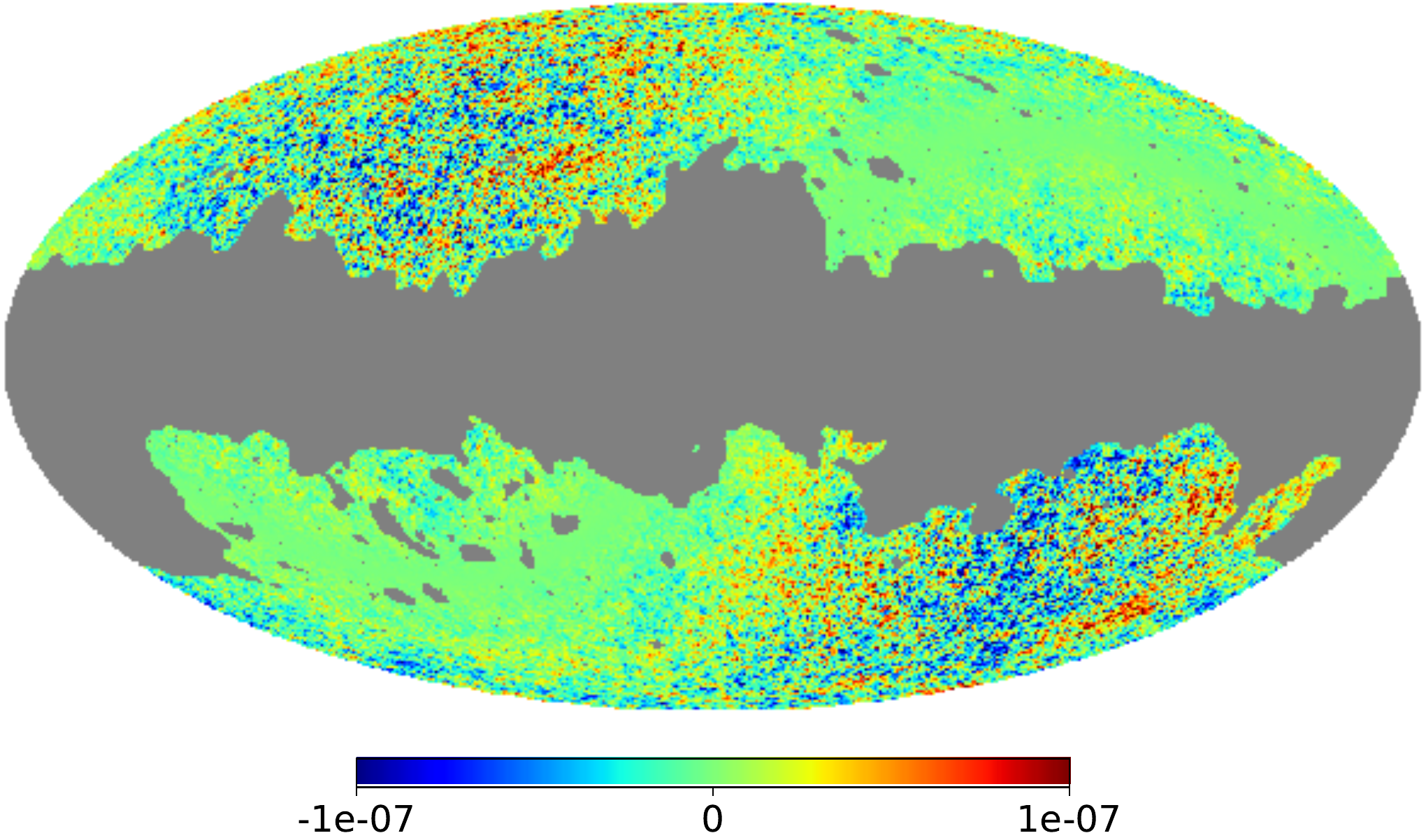}}
\caption{Map of the tSZ effect from the \milca\ component-separation method in $y$-map units (top left) and the expected
modulated CMB signal (top right) generated using the \smicano\ CMB map in units of $T_0$. The bottom left and right panels are the CMB anisotropies modulated in orthogonal directions to the CMB dipole. Note that the map of the tSZ effect (top left) has a different scale bar when compared to the other three (i.e.,\ the modulation
signal is about 50 times weaker).}
\label{fig:template_data}
\end{center}
\end{figure}

In Ref.~\citenum{planck2013-pipaberration} a quadratic estimator was
used to determine the dipole aberration and modulation, in essence using the auto-correlation of the CMB fluctuation
temperature maps.
In this work we use the fact that we know the true CMB
fluctuations with excellent precision and therefore know the signal that should be
present in the $y$~map. Thus, we fully exploit the angular dependence of
the modulation signal and remove much of the cosmic variance that would be
present in the auto-correlation.
In order to implement this idea we define three templates, $B_i$ (with
$i=1,2,3$) as
\begin{align}
  B_i(\hat{\vec{n}}) &= \beta \hat{\vec{n}} \cdot \hat{\vec{m}}_i\, \frac{\delta T}{T_0} (\hat{\vec{n}}),
  \label{eq:template}
\end{align}
where $\beta = v/c$~\citep{planck2016-l01} is $1.23357\times10^{-3}$ and $\hat{\vec{m}}_1, \hat{\vec{m}}_2,
\hat{\vec{m}}_3$ are the CMB dipole direction, an orthogonal direction in the
Galactic plane, and the third remaining orthogonal direction (see Fig.~\ref{fig:template_data}).
Due to the presence of the CMB
dipole, the signal $B_1$ should be present in the $y$~map and so we can directly cross-correlate $B_1$ with our $y$~map to pull out
the signal. Likewise, the cross-correlation of $B_2$ and $B_3$ with our $y$~map
should give results consistent with noise.

Our $y$ simulations are generated by first computing the power spectra of our
data $y$~maps; specifically we apply the {\tt MASTER} method using the {\tt NaMASTER} routine \citep{NaMaster} to account for the applied mask \citep[][]{Hivon2002}. Then we
generate $y$~maps using this power-spectrum with the \healpix\
\citep[][]{gorski2005} routine {\tt synfast}. We finally apply a Gaussian smoothing of
5\arcm\ to model the telescope beam.
For each analysis method we
estimate the amplitude of the dipole ($\hat{\beta}_i$) in each of the three
orthogonal directions. We apply the same analysis procedure on a suite of 1000 $y$
simulations, generated with and without the dipolar modulation term. 

We use two methods of cross-correlation: the
first is performed directly in map-space; and the second is performed in harmonic
space. For both methods we first apply our mask to the templates $B_i$ and the $y$~map. 

In the map-space method we then locate all
peaks (i.e., local maxima or minima) of the template map $B_i$ and select a patch of radius $2\pdeg0$ around
each peak. For every peak we obtain an estimate of $\hat{\beta}_i$ through the simple
operation
\begin{align}
  \hat{\beta}_{i,p} &= \beta \frac{\sum_{k\in D(p)} B_{i,k} y_k}{\sum_{k\in D(p)}
  B_{i,k}^2},
  \label{eq:mapcross}
\end{align}
where $D(p)$ is the collection of all \emph{unmasked} pixels in a $2\pdeg0$
radius centred on pixel $p$, and $p$ is the position of a peak.
Equation~\ref{eq:mapcross} is simply a cross-correlation in map space and by itself
offers a highly noisy (and largely unbiased) estimate.
We then combine all
individual peak estimates with a set of weights ($w_p$) to give our full
estimate:
\begin{align}
  \hat{\beta}_i &= \frac{\sum_p w_{i,p} \hat{\beta}_{i,p}}{\sum_p w_{i,p}}.
  \label{eq:mapweighting}
\end{align}
We choose $w_p$ to be proportional to the square of the dipole, and use weights that
are proportional to the square of the Laplacian at the peak
\citep{Desjacques2008}; this favours sharply defined peaks over shallow ones.
Finally we account for the scan strategy of the \Planck\ mission by weighting
by the 217-GHz hits map\citep{planck2014-a09}, denoted $H^{217}_p$. The weights are then explicitly
\begin{align}
  w_{i,p} &= |\hat{\vec{n}} \cdot \hat{\vec{m}}_i|^2_p \left(\left.\nabla^2(B_i)\right|_p\right)^2 H^{217}_p.
  \label{eq:mweights}
\end{align}
We apply the method for each of our simulated
$y$~maps, in exactly the same way as for the data.

Under the assumption that the $y$~map contains
the template ($B_i$), the $y$ multipoles are Gaussian random numbers with mean
and variance given by
\begin{align}
  s^i_{\ell m} &= \int d\Omega\, \beta\, \hat{\vec{m}}_i\cdot\hat{\vec{n}}\, \frac{\delta T}{T_0} M(\Omega) Y^*_{\ell m},\\
  \sigma^2_{\ell} &= C^y_{\ell} + N^y_{\ell},
  \label{eq:mean_signal}
\end{align}
respectively, where $M(\Omega)$ is the mask over the sphere, $Y_{\ell m}$ are the spherical harmonics, and the $\hat{\vec{m}}_i$ are as defined in Eq.~\ref{eq:template}. We can obtain an estimate of
$\beta_i$ by taking the cross-correlation with inverse-variance weighting.
Our estimator is therefore
\begin{align}
  \hat{\beta}_i &= \beta \sum_{i'} \left[\sum_{\ell m}^{\ell_{\max}}
  s^i_{\ell m}(s^{i'}_{\ell m})^* / \sigma^2_{\ell}\right]^{-1} \sum_{\ell
  m}^{\ell_{\max}} s^{i'}_{\ell m}
  (y_{\ell m})^*/\sigma^2_{\ell}.
  \label{eq:hcross}
\end{align}

\begin{figure}[htbp!]
\begin{center}
\includegraphics[width=\hsize]{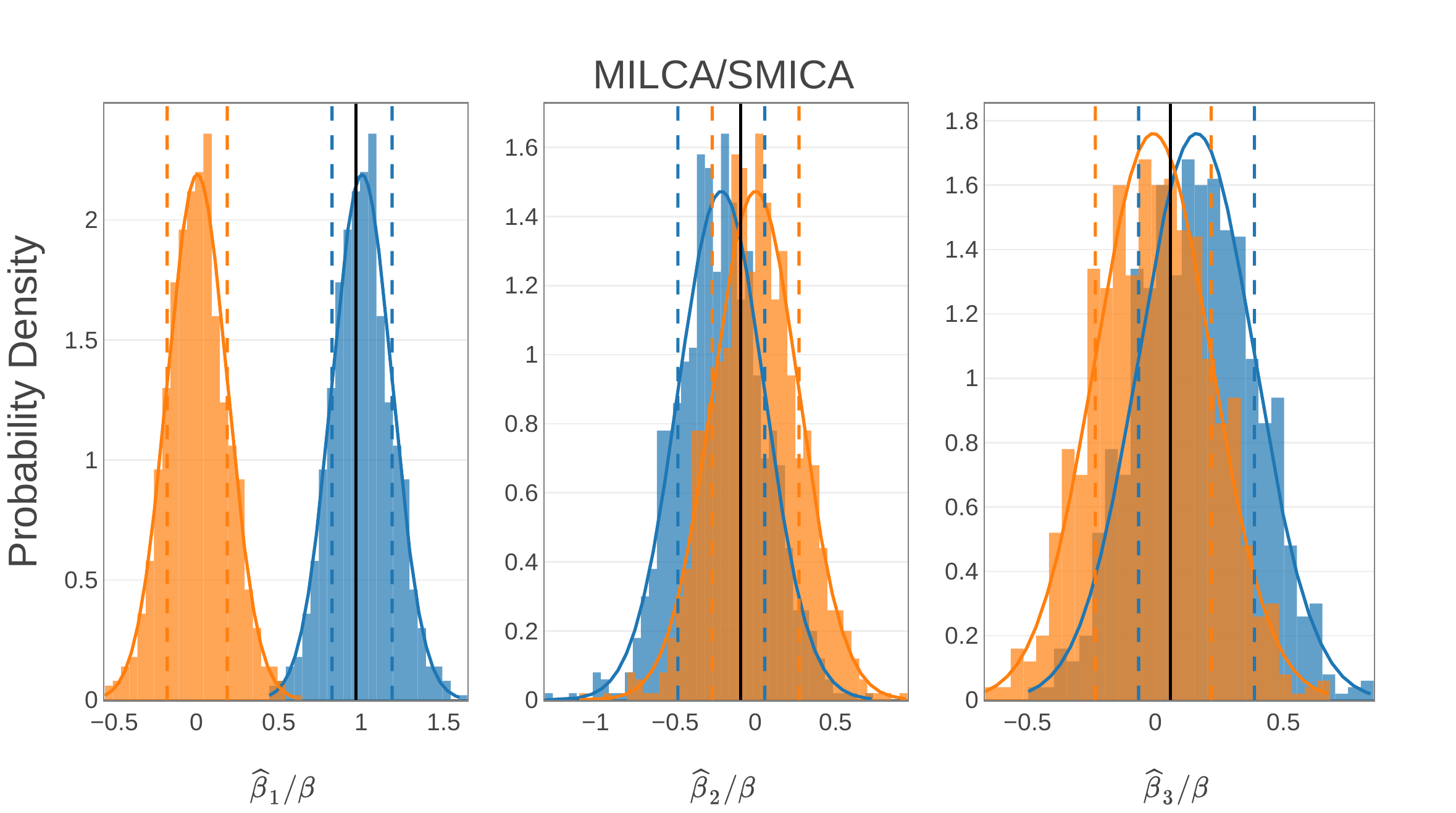}
\caption{Histograms of $\hat{\beta}_i/\beta$ values (with 1, 2, and 3
corresponding to the CMB dipole direction, the Galactic plane, and a third
orthogonal direction) using the map-space analysis for
\milca\ $y$~maps, and for CMB template maps \smicano. Blue histograms are simulations with the dipolar modulation
term, while orange histograms are simulations without modulation. Black vertical lines
denote the values of the real data, demonstrating that they are much more consistent
with the existence of the dipolar modulation term than without it. Dashed lines show the 68\,\% regions for a Gaussian fit to the histograms. To see the full results with all data analysis combinations see Ref.~\citenum{planck2020-LVI}.}
\label{fig:map_histo}
\end{center}
\end{figure}

\begin{figure}[h]
\begin{center}
\includegraphics[width=\hsize]{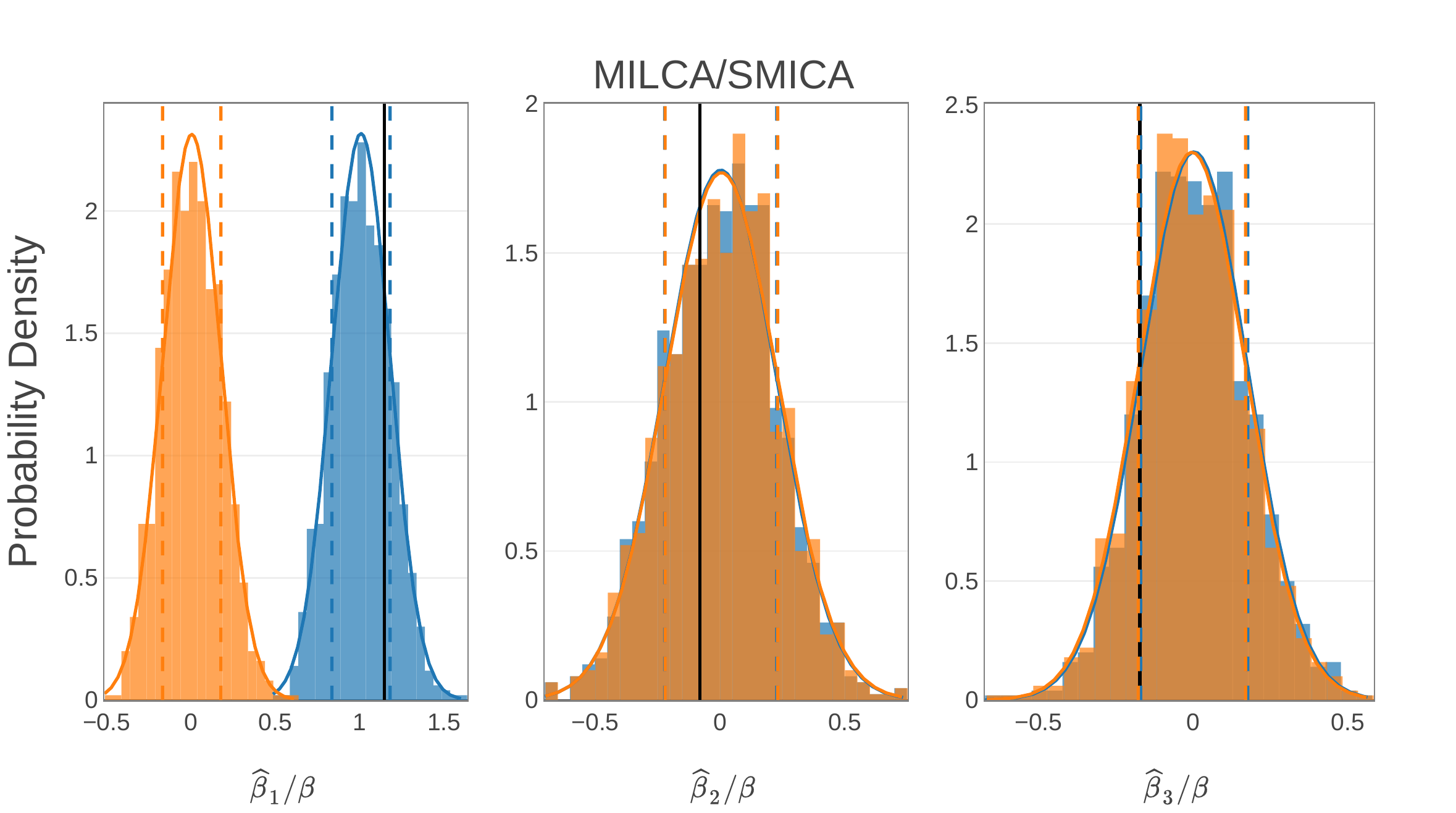}
\caption{As in Fig.~\ref{fig:map_histo}, except now for the harmonic-space analysis.}
\label{fig:har_histo}
\end{center}
\end{figure}

First we compare the consistency of the data with our two sets of simulations
(with and without the dipole term). This comparisons shown in
Figs.~\ref{fig:map_histo} and \ref{fig:har_histo} have blue histograms being the simulations \emph{with} the dipole term and orange histograms being \emph{without}. The data (black line) for \twodilc\
and \milca\ can clearly be seen to be consistent with the
simulations with the dipole term. Further details and analysis may be found in Ref.~\citenum{planck2020-LVI}. 

\section{Conclusion: Distinguishing Intrinsic and Extrinsic CMB Dipoles}
\label{sec:concl}
The frequency-dependent part of the dipolar-modulation signal is agnostic to the source of the large CMB dipole. Therefore, its measurement is an independent determination of the CMB dipole. While it may be tempting to use this signal to distinguish an intrinsic dipole, it has been shown that an intrinsic dipole and a dipole induced by a velocity boost would in fact have the same dipolar-modulation signature on the sky \citep{Challinor2002, Notari2015}

Due to the existence of the CMB dipole, a tSZ map necessarily contains a contaminating signal that is
simply the dipole modulation of the CMB anisotropies. This occurs because CMB
experiments do not directly measure temperature anisotropies, but instead measure
intensity variations that are conventionally converted to temperature variations. This
contamination adds power to the tSZ map in a $Y_{20}$ pattern,
with its axis parallel to the dipole direction. We have measured this effect
and determined a statistically independent value of the CMB dipole, which is
consistent with direct measurements of the dipole. Using a conservative multipole cut on the $y$~map, the significance of the detection of the dipole modulation signal is around 5 or $6\,\sigma$, depending on the precise choice of data set and analysis method.

The question as to whether an intrinsic dipole could ever be observationally distinguished from an extrinsic dipole (i.e. a Doppler boost) remains an open question. The terms discussed in
Eq.~\ref{equ:boostCMB} are based on the assumption of a CMB blackbody spectrum and cannot be
used to distinguish the two, as they would naturally arise whether the CMB dipole
is caused by a boost, or if there is for some other reason a dipole on the sky with the same
magnitude and direction.

\section*{Acknowledgements}
We would like to acknowledge the support of the Natural Sciences and
Engineering Research Council of Canada. Some of the results in this
paper have been derived using the {\tt HEALPix} package. Results are
based on observations obtained with Planck
(http://www.esa.int/Planck), an ESA science mission with instruments
and contributions directly funded by ESA Member States, NASA, and
Canada. We would also like to thank Dagoberto Contreras for
collaboration on topics within this paper.
 
\bibliographystyle{ws-procs961x669}
\bibliography{dipolebib}

\end{document}